# Application of Optical Tweezers in the Study of Emulsions for Multiple Applications

*Qifei Ma[a], Huaizhou Jin[b]\*, Xiaoxiao Shang[a], Tamas Pardy[c,d], Ott Scheler[d], Simona Bartkova[d], Dan Cojoc[e]\*，Denis Garoli[a,f,g]\*, Shangzhong Jin[a]\**

[a] *College of Optical and Electronic Technology, China Jiliang University, Hangzhou 310018, China*

[b] *Key Laboratory of Quantum Precision Measurement, College of Physics, Zhejiang University of Technology, Hangzhou, China*

[c] *Thomas Johann Seebeck Department of Electronics, Tallinn University of Technology, Estonia*

[d] *Department of Chemistry and Biotechnology, Tallinn University of Technology, Estonia*

[e] *CNR-Istituto Officina dei Materiali (CNR-IOM),SS 14 km 163.5, Area Science Park Basovizza,34149, Trieste, Italy*

[f] *Istituto Italiano di Tecnologia, Via Morego 30, 16136 Genova, Italy*

[g] *Università degli studi di Modena e Reggio Emilia, Dipartimento di Scienze e Metodi dell'ingegneria, Via Amendola 2, Reggio Emilia, Italy*

*\* Corresponding author. E-mail address: jinsz@cjlu.edu.cn*

Emulsions are ubiquitous in everyday life and find applications in various industries. Optical tweezers (OTs) have emerged as the preferred method for studying emulsion dynamics. In this review, we first introduce the theory of optical trapping and emulsion stability. We then survey applications in the manipulation of emulsions, stability mechanism, the processes of aggregation and coalescence, and important responsive and switchable behaviors. And we overview the instrumentation framework of various OT setups, and evaluate their complexity and cost with a view towards the democratization of this technology. Following this, we delve into basic experimentation methods, the challenges associated with using OTs in emulsion applications. Additionally, we present a promising research outlook, including studies on stability mechanism of emulsions stabilized by compound/mixed emulsifiers or rigid/soft particles, as well as dynamic processes of responsive or functional emulsions.

**Keywords:**
Optical tweezer; Emulsion; Stability; Aggregation; Coalescence; Switchable behavior

**Highlights:**
1) Analyzing the principles of manipulation and measurement of emulsions using

optical tweezer expounds the feasibility and necessity of the study.

2) Classifying the applications of emulsion using optical tweezer, including manipulation of emulsions, stability, the processes of aggregation and coalescence, and responsive and switchable behaviors, shows a rapidly rising interest in emulsions and optical tweezers.

3) Summarizing instrumentation framework of setups and basic experimentation methods guides the further study of emulsions using optical tweezers.

**Introduction**

Emulsions are multiphase unstable systems consisting of two liquids that typically do not mix, along with an emulsifier that facilitates their mixing. One liquid forms the continuous phase, while the other is dispersed as small droplets. Essentially, there are three types of emulsions: oil dispersed in water (oil-in-water, O/W), water dispersed in oil (water-in-oil, W/O), and a more complex type known as multiple emulsions that can be prepared as water-in-oil-in-water (W/O/W) emulsions[1, 2].

Emulsions have diverse applications across industries, including food and beverage production[3], pharmaceutical formulations[4] (e.g., drug delivery systems), and cosmetic products[5] (e.g., creams, lotions, makeup). In biomedical research different emulsion systems are being applied as nanoscale (bio)chemical reaction compartments for diagnostics[6] and high-throughput screening of novel biomolecules[7]. These applications highlight the importance of emulsions in various sectors, where they contribute to product quality, performance, and functionality.

Research aims to optimize emulsion properties for specific applications and explore novel formulations. Therefore, there are many interesting research studies about emulsions, identifying factors influencing stability, studying the kinetics of droplet coalescence, researchers explore functional emulsion response processes, such as controlling release kinetics, improving bioavailability, and targeting delivery to specific sites within the body. It is fundamental to have techniques to study behaviors and mechanisms of emulsions at the microscale, which can achieve single-droplet analysis, real-time observation, and real-time monitoring of processes in functional emulsions. Nowadays, techniques such as atomic force microscopy (AFM), magnetic tweezers (MTs), and OTs allow for the measurement of forces and displacements induced at the single-cell and single-molecule levels. These techniques have been employed to improve the experimental design for emulsion studies[8, 9] offering nanometer and sub-millisecond spatial and temporal resolution. The force is determined by measuring the displacement of a probe (such as a dielectric or magnetic microbead, or a cantilever tip), which is characterized by an elastic constant (stiffness). With stiffnesses typically in the range $10 - 10^5$ pN nm$^{-1}$, AFM allows for the measurement of forces typically in the range $10 - 10^4$ pN. MT and OT probes typically have stiffnesses in the range $5 \times 10^{-3} - 1$ pN nm$^{-1}$, making them suitable for measuring lower forces than AFM, typically in the range 0.1 - 200 pN. The manipulation, imaging, and measurement of forces by AFM exceeding 100 pN have primarily focused on droplets with diameters ranging from 20 to 200 μm, providing insights into the deformation process upon interaction with substrates or other droplets and surface tension[10]. MT are commonly employed to measure low forces. However, in combination with microfluidics, microMT capable of generating μN forces have been developed to trap and extract magnetic particles from droplets, enabling physical separation in single cell-based droplets[11].

Complementary to AFM and MT, OT allows for the manipulation and force measurement of dielectric particles without mechanical contact, making it the method of choice for studying many emulsion droplets with below 10 μm in diameter. OT enables the suspension of emulsion droplets in specified positions within the liquid, allowing for strict control over environmental conditions[12-15]. Additionally, OT can be combined with various optical techniques on the same microscopy platform, such as fluorescence microscopy[16-18], enabling real-time tracking of droplet position and dynamics. OT can also be combined with Raman spectroscopy[19, 20], facilitating the trapping of single emulsion droplets and the collection of molecular information regarding chemical-physical properties and interactions with the environment.

However, there are numerous challenges associated with using OTs in emulsion applications. The first challenge is the low-throughput nature of the experiments and the relatively limited statistical data derived from them. Another challenge relates to the various instrumental configurations, experimental settings, and data analysis pipelines developed in different laboratories, which make straightforward comparisons between different studies difficult. Manipulation speeds are another important consideration, particularly for constant force measurements. To fully exploit the potential of OTs as a tool for emulsion applications, researchers are exploring different approaches, which will be discussed here. In this review, our aim is to explore the primary applications of OTs in emulsion dynamics. Our discussion will focus on the following areas: 1) manipulation of emulsion droplets; 2) control aggregation and coalescence of emulsion droplets; and 3) online monitoring and measurement of switching behaviors of surface-active colloid spheres and responsive emulsions. By exploring these applications of OTs in the study of emulsions, we aim to provide a comprehensive overview of the technique's potential in this field.

**Principle of optical trapping and force measurement by OT**

The trapping and manipulation of dielectric microparticles using a single-beam gradient force optical trap was demonstrated by Ashkin and his group in 1986[21] paving the way for new applications in physics, chemistry, nano-biotechnology, and biophysics. In recognition of his exceptional contribution, Ashkin was awarded the Nobel Prize in Physics 2018, "for the OTs and their application to biological systems". The key of the single-beam gradient laser trap is the utilization of a microscope lens with a high numerical aperture (NA>1). This lens allows the laser beam to be tightly focused, enabling the creation of a three-dimensional (3D) optical trap located near the focus of the lens, and thus extending the 2D trapping achieved by a single beam focused by a low NA lens (Figure 1A).

Using the ray optics approach, the light beam is decomposed into individual rays, each with its own intensity and polarization, propagating in straight lines through media with uniform refractive index. Each ray can change direction and polarization through reflection and refraction at dielectric interfaces, following Fresnel formulas. Considering a single ray of power P hitting the dielectric sphere at an angle of incidence θ with incident momentum per second of $n_1P/c$ (where $n_1$ is the refractive index of the medium, and c is the speed of light), the force components[22] are given by:

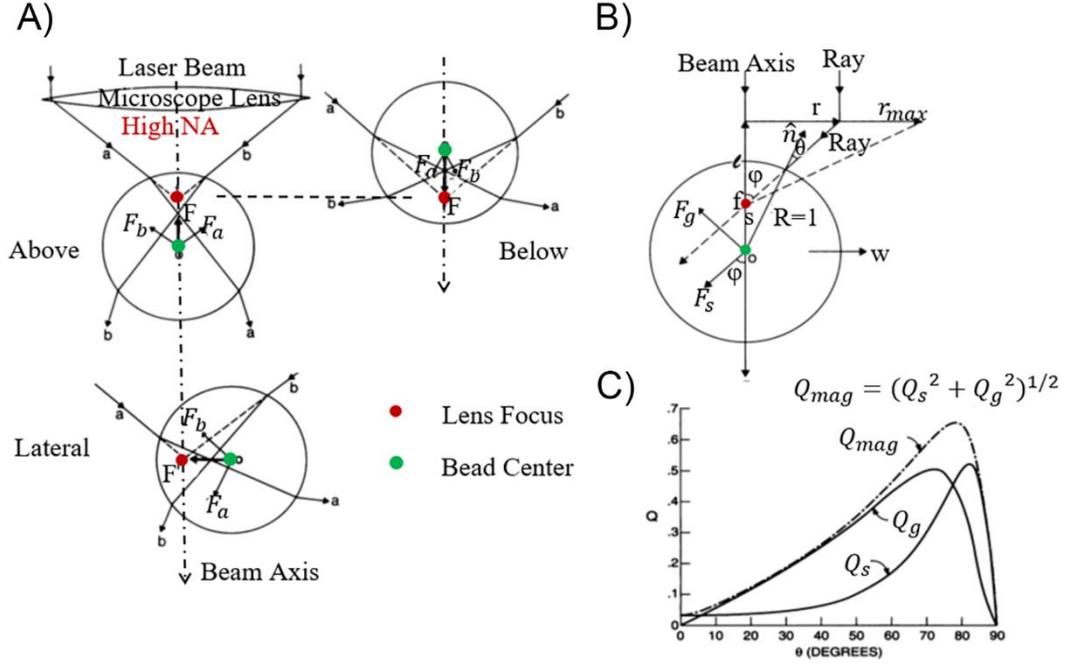

**Figure 1.** illustrates the optical trapping model using ray optics. (A) Trapping of a dielectric microbead: the bead above the focus is pulled toward the lens focus; the bead below is pushed toward the focus, and the lateral bead is attracted toward the focus. (B) Gradient and scattering force components arising from the refraction of an incident light ray. (C) Magnitude of the gradient and scattering force for a single ray, for a polystyrene bead in water, as a function the incidence angle. The magnitude of the resulting force $F_{mag} \sim Q_{mag}$ (Reproduced with permission from Ref [22]). The optical trapping and manipulation of dielectric particles of spherical shape and size larger than the wavelength of light can be readily explained using ray optics and the conservation of momentum of light(Reproduced with permission from Ref [22, 23]).

$$F_Z = F_S = n_1P\{1 + R\cos\cos 2\theta - T^2[\cos\cos(2\theta - 2r) + R\cos\cos 2\theta]/(1 + R^2 + 2R\cos\cos 2r)\}/c$$

(1)

$$F_Y = F_S = n_1P\{R\sin\sin 2\theta - T^2[\sin\sin(2\theta - 2r) + R\sin\sin 2\theta]/(1 + R^2 + 2R\cos\cos 2r)\}/c$$

(2)

Where θ and r represent the angles of incidence and refraction, respectively, while R and T denote the Fresnel reflection and transmission coefficients (Figure 1B). The force component $F_z$, pointing in the direction of the incident ray is called the scattering force $F_s$ for a single ray, while the $F_y$ component pointing in a direction perpendicular to the ray is denoted as the gradient force $F_g$. For a laser beam, the scattering and gradient forces are defined by the vectorial sums of the scattering and gradient force contributions of the individual rays comprising the beam. The magnitude of the optical force is:

$$F = Qn_1P/c$$

Where

$$Q = \sqrt{Q_s^2 + Q_g^2}$$

（3）

The dimensionless coefficient Q takes into account the material and shape of the particle and can have a value of maximum Q=2.

Due to thermal motion, the position of the optically trapped particle fluctuates around the point of equilibrium, where the light intensity is maximum[21]. For small distances from this point (typically 0 - 400 nm), the trap potential can be described as a harmonic/parabolic potential (Figure 2A), in which the trapped particle tends to reach the potential minimum. For a parabolic potential, the restoring force exerted on the particle is proportional to the position x, by an elastic constant k, called trap stiffness (Figure 2B), the optical trap behaving as a Hookean spring:

$$F = -k_{trap}x \qquad (4)$$

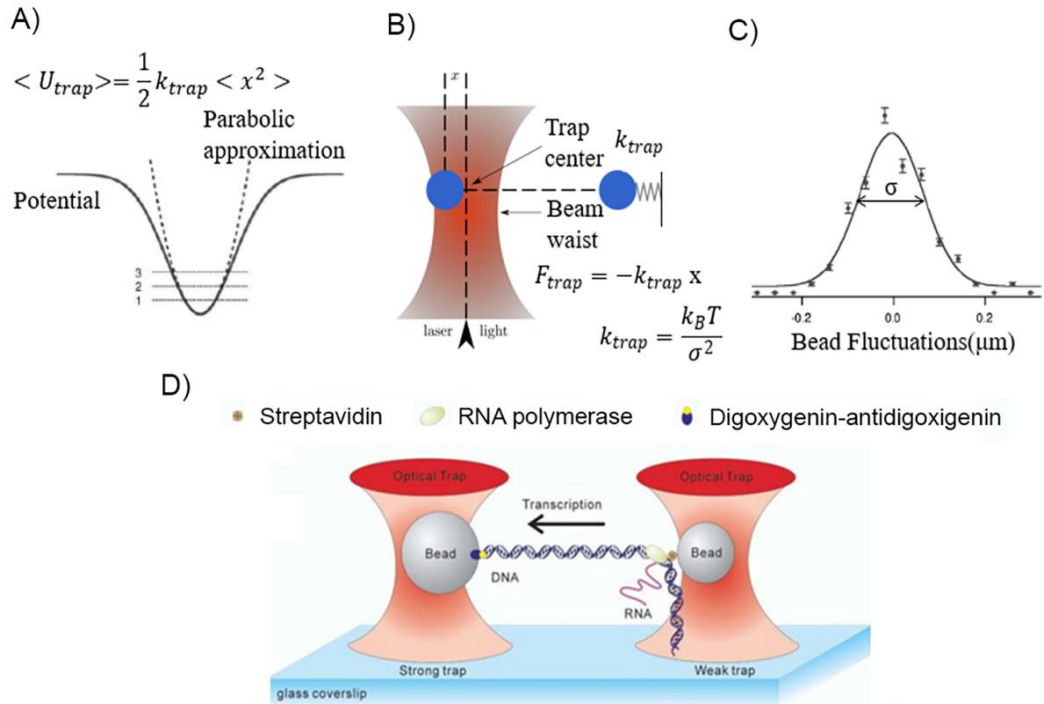

**Figure 2.** (A) Optical trap potential. (B) Optical trap as a Hookean spring. (C) Distribution of the bead position. (D) and example of a dual beam optical trap for DNA transcription experiment(Reproduced with permission from Ref[24]).

The value of the trap stiffness can be determined by tracking the position of the bead in the trap for 3 - 5 seconds at a frequency of about 5 kHz and using the Boltzmann distribution (Figure 2C) or the Equipartition theorem :

$$k_{trap} = k_B T/\sigma^2$$

(5)

Where $\sigma^2 = \langle x^2 \rangle$ is the variance, $k_B$ the Boltzmann constant and T the temperature.

Aside of this method, there are also other techniques to calibrate the trap: passive and active power spectrum, drag force, light momentum change[25, 26].

A useful device for the study of emulsion droplets is the dual-beam OTs. This system employs two optical traps, allowing the trapping and independent manipulation of two different particles in liquid[24]. While the primary biological application of dual-beam OT is measuring the biophysical properties of molecular motors, it can also be applied to droplets. In this setup, two trapped droplets can be brought together to interact, and the displacement of each droplet is measured separately. One advantage of this configuration is its higher spatial resolution compared to a single trap, which can be achieved even with droplets of larger diameters (> 1μm).

**Theory of emulsion stability**

The interactions at micro scales of chemical reagents, solid particles, bubbles, droplets, and solid surfaces in complex fluids play a crucial role, which impact the macroscopic performance and efficiency of related engineering processes. Classical intermolecular and surface interactions include Derjaguin-Landau-Verwey-Overbeek (DLVO) interactions (i.e., van der Waals (VDW) and electric double layer (EDL) interactions) and non-DLVO interactions such as steric and hydrophobic interactions[27].

*DLVO theory*

The DLVO theory[28] has been widely applied to describe the stability of colloidal spheres in aqueous solution, which includes both VDW and EDL interactions.

VDW forces are ubiquitous between all molecules and surfaces, because VDW interactions result from the associated fluctuating electric dipole moments when two molecules approach each other. The VDW force[29] can be calculated using Eq(6).

$$F_{VDW}(h) = -A_H R/6h^2$$

(6)

Where $A_H$ is the Hamaker constant between two spheres (with radius $R$) in an aqueous solution, and $h$ is the surface-to surface distance between two spheres.

An EDL is generated when a solution containing ions is in contact with a charged surface. In case of spherical particles, if spheres carry the same charge, the EDL surface interactions are repulsive, preventing the aggregation and precipitation of the spheres. It is more convenient to calculate the EDL force[30, 31] with the surface distance of spheres directly, which can be demonstrated using Eq(7)

$$F_{EDL}(h) = (e^2 Z^2 / 8\pi\varepsilon_0 \varepsilon_r R^3) \kappa^{-1} exp(-\kappa h)$$

(7)

Where $Z$ is a process variable, $R$ is the radius of spheres, $\varepsilon_0$ is the permittivity of vacuum, $\varepsilon_r$ is the relative permittivity of the aqueous phase solution, $e$ is the elementary charge, and $\kappa^{-1}$ is the Debye length. The Debye length can be calculated by Eq(8)

$$\kappa^{-1} = (\varepsilon_0 \varepsilon_r k_B T)/(2 \times 10^3 N A e^2 I)$$

(8)

Where $k_B$ is the Boltzmann constant, $T$ is the thermodynamic temperature, and $I$ is the intensity of ions in aqueous solution. $I$ can be calculated by Eq(9)

$$I = \frac{1}{2} \sum_i c_i z_i^2$$

(9)

Where $c_i$ is the concentration of ions, $z_i$ is the electric charge of ions. $Z$ can be calculated by Eq(10) and Eq(11)

$$Z = e\xi R/(k_B T \lambda_b)(1 + \kappa R)$$

(10)

$$\lambda_b = e^2/(4\pi\varepsilon_0 \varepsilon_r k_B T)$$

(11)

Where $\xi$ is the $\xi$-potential of spheres and $\lambda_b$ is the Bjerrum length.

Hence, the total interaction force between two spheres can be calculated using Eq(12).

$$F_{tol}(h) = F_{VDW}(h) + F_{EDL}(h)$$

(12)

*Non-DLVO theory*

Besides DLVO interactions, there are several other interactions, referred as non-DLVO interactions, including steric force, depletion force, polymer bridging interaction, hydrophobic effects, and hydration force, which can also impact the interactions of particles.

- Steric force[32] arises from the compression of polymer chains when two particles stabilized by polymers come into close proximity.
- Depletion force. For spheres stabilized by polymers, if the polymer that does not adsorb or weakly adsorb onto any surfaces of spheres during the approach of two spheres, the polymer between the spheres is squeezed out, leaving a bare surface, thus a "depletion zone" will appear. There is a difference in polymer concentration between the depletion zone and polymer solution, resulting in a difference in osmotic pressure, which causes water molecules to migrate from the depletion zone into the bulk solution, creating a depletion force[33, 34].

- Bridging force. For spheres stabilized by polymers, if relatively low amounts of polymers adsorb onto the surface of one sphere, the other ends of these polymers may bind to other spheres, resulting in adhesive bridging force[35, 36] between the two surfaces.

- Hydrophobic interactions. Hydrophobic molecules are non-polar molecules, typically possessing long carbon chains that cause them to self-associate in aqueous solutions. Hydrophobic interactions[37] can drive the aggregation of hydrophobic moieties in water mediums and adjust molecular or biomolecular conformation on a macro scale, as well as facilitate oil-water separation, which usually exist between different proteins and other biochemical molecules. For example, when proteins fold in water, they tend to bury hydrophilic groups and expose hydrophilic groups.

**Applications**

In this section, applications of OTs with relation to emulsions will be comprehensively reviewed. OTs, a powerful tool in the field of biophysics and nanotechnology, have found diverse applications in manipulating and studying colloidal systems such as emulsions. To gauge the interest of OTs in various sub-fields, we present an overview of the number of publications indexed in Scopus, highlighting the growing interest and research contributions in this evolving area of study.

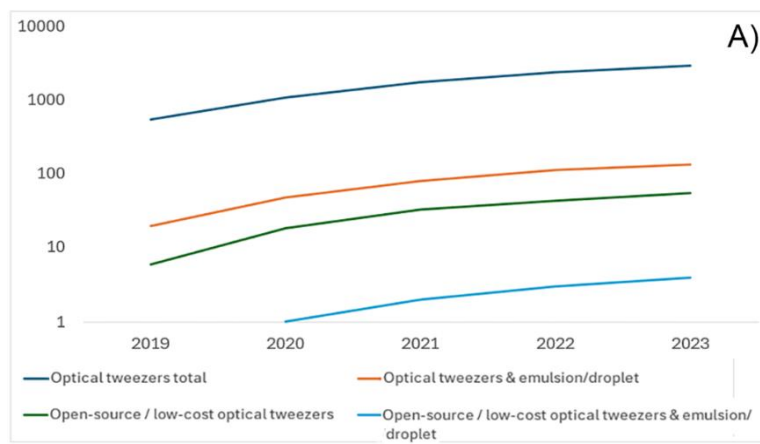

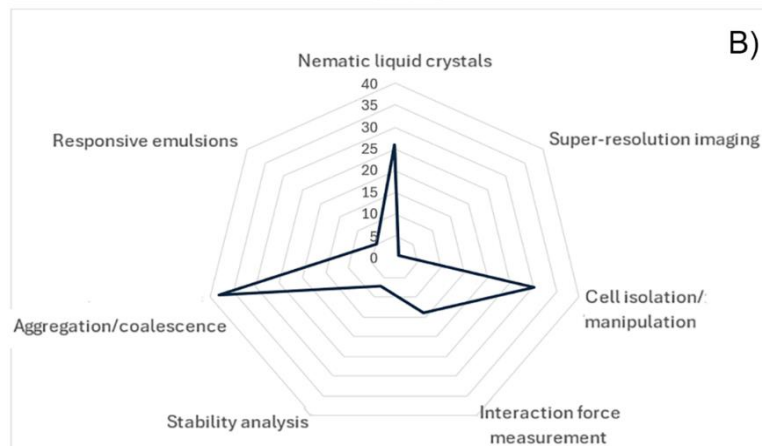

**Figure 3**: According to the Scopus publication index, there has been (A) a steadily rising interest in optical tweezers in general, and a rapidly rising interest in emulsions/droplets within the last 5 years. Certain application areas (B), such as droplet aggregation/coalescence, nematic liquid crystals and cell isolation/manipulation (in/with droplets) stand out in particular and will be discussed further in this review.

*Manipulation of emulsion droplets*

Manipulation of materials at the microscale is vital for evaluating the microscopic properties of soft materials and biomaterials. OTs play an important role in precise control of micro-objects, such as colloids, micro-organisms, and cells.

A first interesting example is the optically driven nematic liquid crystals (NLC) droplet rotator. It offers a powerful method for manipulating droplets with high precision and control. By utilizing the unique properties of NLC and advanced optical techniques, these devices have the potential to revolutionize applications in microfluidics, optics, and biological research. Suwannasopon et al. [38] provided a way to rotate NLC droplets by the combination of metalens and OTs in 2019. Ultra-thin metalenses have great potential to replace traditional lenses in the future, which can be used in lab-on-a-chip systems. The liquid crystal is well combined with this miniature OTs device, creating an ideal optical motor. They can be used in several microscopic systems for the motion and flow control. Figure 4A and 4B show the structure of metalens. Figure 4C is a schematic diagram of sample chamber attached on top with metalens. Saito et al. [39] performed an optically driven NLC droplet rotator in 2022. In this study, they combined elliptically polarized laser light with OTs to control droplet rotation, also analyzing the rotation mechanism on the basis of the arrangement of liquid crystal molecules within the droplets. For NLC droplet, the waveplate effects and light-scattering process lead to rotation. For cholesteric LC(ChLC) droplets, the waveplate effects and Bragg reflection elaborate the rotational behaviour. Figure 5A and 5B show the microscopy images of NLC droplets and ChLC droplets.

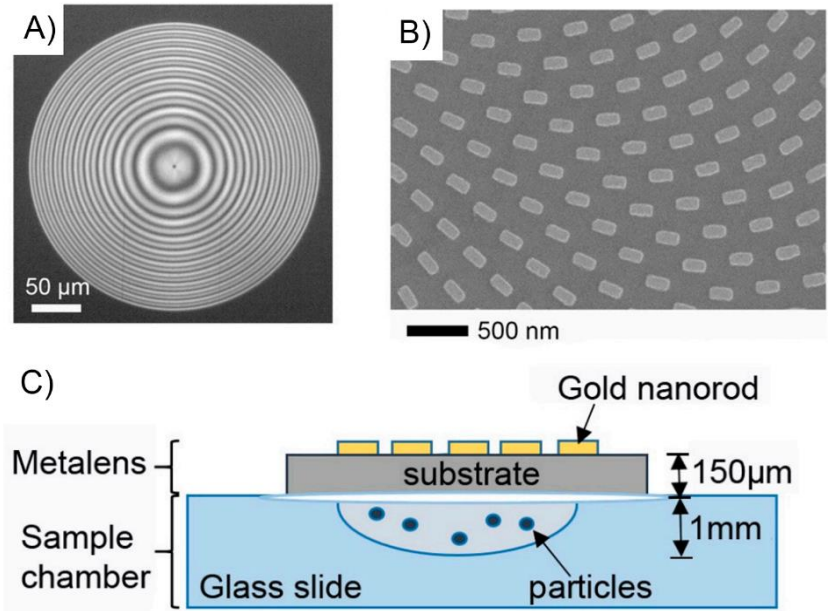

**Figure 4** (A) Optical microscopy image of the metalens surface. (B) Scanning electron microscopy(SEM) image of the lens. (C) Schematic diagram of sample chamber attached on top with metalens (Reproduced with permission from Ref.[38]).

OT and droplets find important applications also in imaging. In order to overcome the diffraction limit and magnify nanostructure imaging, microsphere-assisted imaging[40] has become an irreplaceable tool in life sciences and precision measurement due to its advantages of economy and label-free operation. Wen et al.[41] and Lin el al.[42] combined the OT and microsphere to overcome the disadvantages of small size and limited imaging range of conventional solid microspheres when imaging large sample areas. However, most microspheres have low biocompatibility. Therefore, the use of a single biological element as a photonic element with well-defined characteristics becomes an interesting new paradigm for biophotonics research. In particular, Chen et al.[43] found that lipid droplets in mature adipose cells can be used as completely biocompatible microlenses to improve microscopic imaging and detecting signals inside and outside cells. Additionally, they used OTs to control lipid droplets to locate targets and perform real-time imaging inside cells.

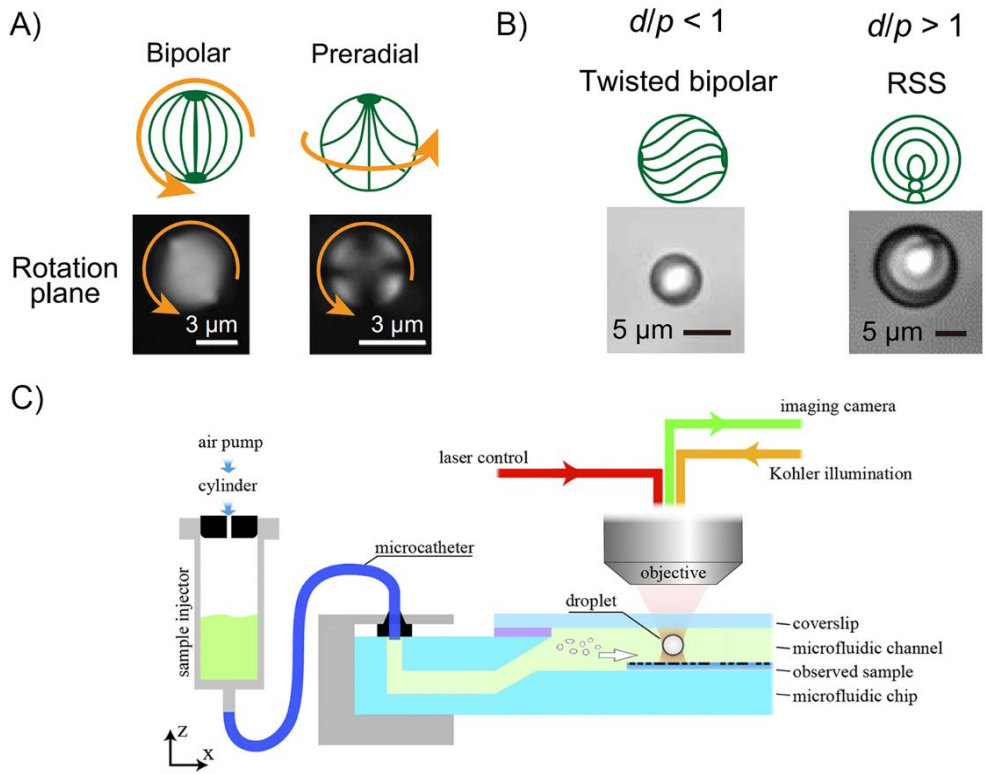

**Figure 5** (A) polarizing microscopic images of NLC droplet. For larger droplet, the inner structure is bipolar(left), for smaller droplet, inner structure is preradial(right). The arrows indicate the rotation directions (Reproduced with permission from Ref.[39]). (B) Bright field microscopic images of ChLC droplets with twisted bipolar structure and RSS (Reproduced with permission from Ref.[39]). (C) OT-based microdroplet generation scheme(Reproduced with permission from Ref.[44]).

Zhai et al[44] proposed a novel microdroplet-assisted imaging technology based on OTs and microfluidic chip. By using the properties of photogeneration, the OT system can produce droplets of any size and achieve different sizes of FOV for magnification imaging. Compared with solid microspheres, droplets still cannot possess the same super-resolution imaging capability, but they have the characteristics of controllable generation technology, larger imaging field of view, which may provide new ideas for the development of microsphere-assisted imaging technology. Additionally, by changing the droplet viscosity or the original solution, the magnification of droplet imaging can be further improved. The microdroplet generation scheme based on OT is shown in Figure 5C. The relative position between the sample chamber and the objective can be adjusted at micrometre scale by a displacement stage.

One potential future direction for enhancing manipulation of microdroplets via OTs is use of open microfluidic channel systems[45] such as by Khor et al., which allows direct manipulation of droplets with PTFE - coated tweezers. The coating prevents strong adhesion of aqueous droplets to the tweezers, thus also facilitating droplet release. The PTFE-coated tweezers enables both lateral and vertical (picking up droplets) droplet transport in the microfluidic system. In the future, this approach could be used to sort droplets based on a reaction outcome or transfer droplets to a different part of a chip for further usage, all of which could be beneficial for many fields such as chemical analysis, drug delivery, and biological research.

Another potential lies in employing acoustic instead of optical tweezers. In their study, Lin et al. successfully used single-beam (focused beam/vortex beam) acoustic tweezers[46] as a selective sparse sampling method for water-in-oil droplets. They generated the droplets with fluorescein dye (for later visualization and analysis), by using of a microfluidic device with flow-focusing junction and droplet sizes ranged from 20 to 150 μm. Droplets smaller than half a wavelength could be trapped by acoustic vortices, while larger water droplets could be trapped via focused acoustic beams. This enabled targeting and extracting selected droplet microreactors based on their size in the microfluidic system and analyzing their content. There is great future potential of using acoustic tweezers for droplet manipulation in combination with microfluidics in many different fields. This setup is especially useful for enhancing high-throughput drug screening assays, fluorescent labeling is not required for sorting and the handling of droplets is gentle.

*Stability of emulsions*

The stability of emulsions is essential for many industrial productions. The understanding of emulsion stabilization mechanism relies on the understanding of interaction forces between single droplet coated stabilizers.

OTs can be used to understand fundamental colloidal properties and their role in emulsion stabilization. The electrostatic interactions between highly charged particles dispersed in electrolytes have been obtained. Among the others, Crocker et al [47, 48] used OTs to measure the pair potential. A variety of measurements revealed purely repulsive interactions, which were quantitatively consistent with predictions of the DLVO theory. Elmahdy et al [49] studied the forces within single pairs of charged colloids in aqueous solutions of ionic liquids using OTs. The force curves were described by a size-corrected screened Coulomb interaction method. They obtained effective surface charge density from force curves, which altered with concentration and PH.

Additionally, OTs can be used to measure the elastic properties of polymers, Gutsche et al [50] summarized a review about miro-rheology on (polymer-grafted)colloids using OTs. They presented several novel miro-rheological and microfluidic experiments, discussed force measurements and nonlinear responses within single pairs of DNA-grafted colloids, and even analyzed the drag force on colloids pulled through a polymer solution. Mahdy et al [49] and Dominguez-Espinosa et al[51] have used OTs to measure a steric interaction of less than 20 pN between polymer brushes and clarified the entropic (osmotic) contribution of the counterions in the brush layers. Murakami et al [52] examined the long-range electrostatic interaction between polyelectrolyte brush surfaces directly using OTs.

Depletion interactions also play a significant role due to their influence on the behavior and stability of colloidal systems. Therefore, it was vital to quantitatively measure depletion interaction and predict the stability of colloids in advance. Liu et al[53] quantitatively measured the interaction forces (including depletion) between two silica particles induced by an ionic surfactant sodium dodecyl benzene sulfonate (SDBS) using OTs. Figure 6A shows force vs separation for a couple of silica particles during both the approach and retraction processes phases with SDBS micelles. Force hysteresis occurred in the retraction portion of the curve when the concentration of SDBS is larger than critical micelle concentration (CMC). The interaction forces between the same couple of silica particles with and without SDBS were measured in situ. By subtracting the force curves between the same couple of silica particles in the absence of SDBS from the force curves in the presence of SDBS, the pure depletion force can be quantitatively calculated. Figure 6B shows the calculated depletion forces.

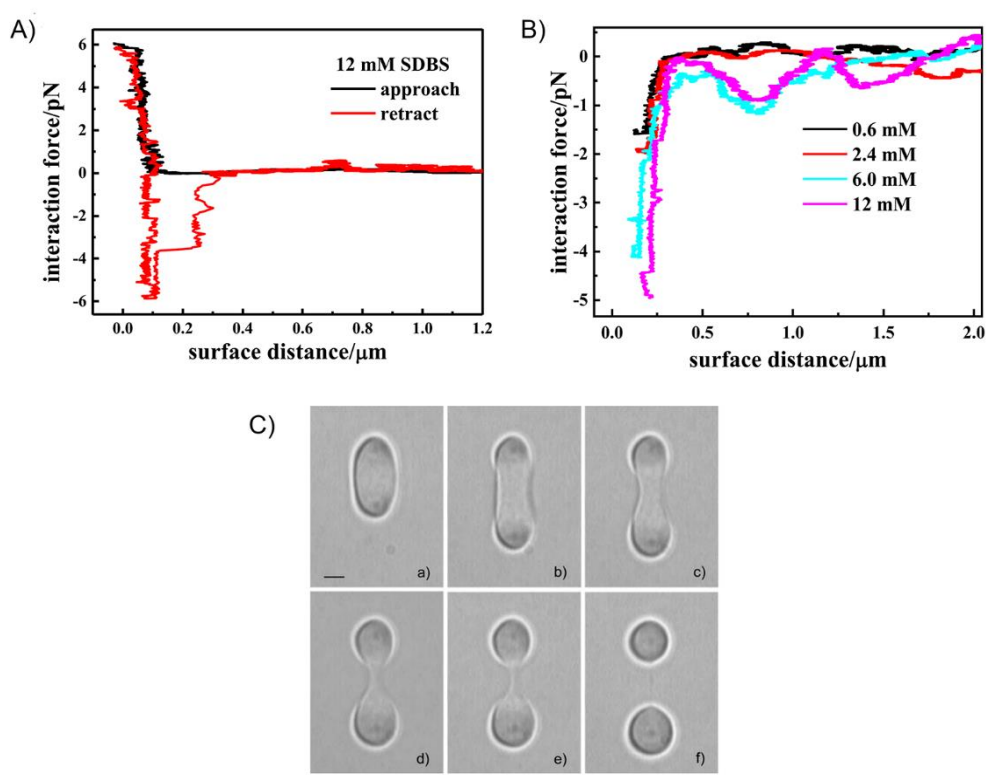

**Figure 6** (A) Force vs separation for one single pair of silica particles SDBS at 12 mM, the approach (black line) and retraction (red line) (Reproduced with permission from Ref [53]). (B) Calculation depletion forces between silica particles with the presence of different concentrations of SDBS(Reproduced with permission from Ref [53]). (C) With very low tensions the droplets form dumbbell shapes on extension and can be separated and rejoined (Reproduced with permission from Ref[14]).

Several theoretical models have been developed to analyze interactions between colloids, which can serve as a basis for studying emulsion interactions and stability. Ward et al[14] reported an optical deformation technique for micron-sized O/W emulsions with ultralow interfacial tensions, achieved by the manipulation of multiple optical trapping sites within the droplets. They used this technique to characterize interfacial properties in the emulsion. The results were in good agreement with those obtained by Aveyard et al [54] and Mitani et al [55] using spinning drop interfacial tensiometry. Figure 6C shows multiple deformed droplets.

In 2014, Julie et al [56] qualitatively measured interactions between emulsion droplets for the first time. They compared force curves between emulsion droplets stabilized by micro- and macromolecular emulsifiers and explored the effects on depletion interaction. They observed the phenomenon that the biopolymer layer of sugar beet pectin (SBP) covering the emulsions surface reorganized during compression. Figure 7A shows the force versus time curve between two emulsion droplets stabilized by highly methylated SBP. There is a steady increase in repulsive force as the droplets approach due to the initial overlap of electric bilayer. At a certain point in the approach segment of the curve, the force suddenly drops to a lower level. After retraction, the force re-established at the same maximum level. The force reduction is reversible upon retraction, and coalescence of the droplets does not occur, indicating the rearrangement of the polymer layer. The force curves from OTs display the dynamics of macromolecular emulsifier layer. Additionally, attractive van der Waals forces can be measured during non-deforming polystyrene beads measurements, as shown in Figure 7B. Although they didn't quantitatively analyze the force curve, these results are promising and imply that OTs can be a useful tool for researchers in the exploration of emulsion droplet interactions and stability.

Salt is known to affect the electrostatic force between two colloidal particles. An increase in salt concentration will reduce the Debye Length, leading to shorter distances of repulsion, which usually causes instability of the colloidal solution. Griffiths et al [12] proposed a novel method to measure local salt concentration using OTs based on this principle. A single couple of particles or emulsion droplets were kept in a microfluidic channel close to an interface formed between milliQ water and a 5 mM NaCl solution. As ions gradually diffused away from the interface, the salt concentration gradually altered, causing the force- separation curve to shift over time, as shown in Figure 7C. The Debye length can be fitted from the force-separation curves, and the local salt concentration can be obtained using Eq(13), as demonstrated in Figure 7D, which was consistent with a relevant diffusion equation.

$$\kappa^{-1} = 0.304/c_{bulk}^{0.5}$$

(13)

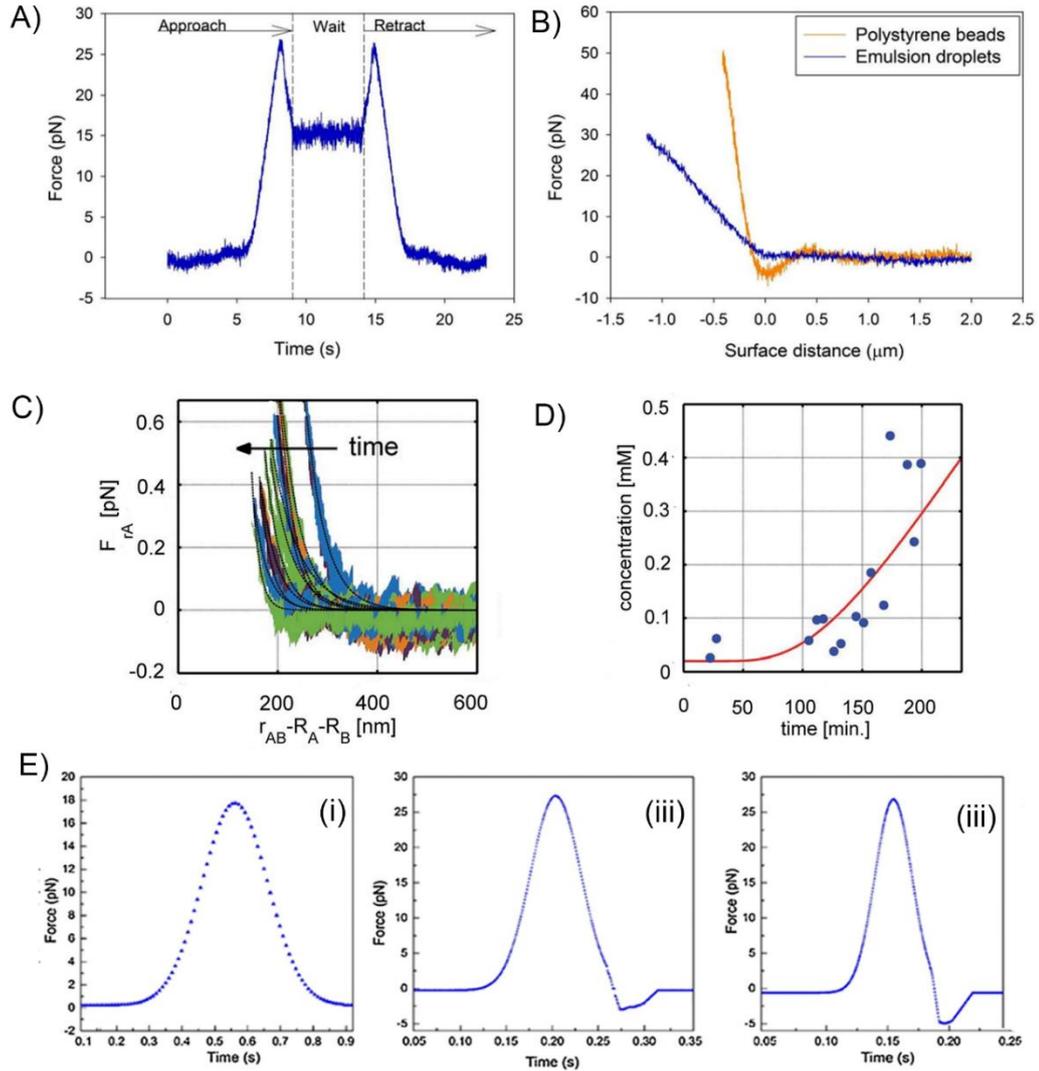

**Figure 7** (A) Approach and retract curve for SBP stabilized droplets in MQ water (Reproduced with permission from Ref.[56]). (B)Approach curves for compression of a pair of rapid beads (2.7μm) and a pair of emulsion droplets (2.5μm) in MQ water(Reproduced with permission from Ref.[56]). (C) Force–separation curves of emulsion droplets as a function of time as the local salt concentration was increasing owing to the diffusion of ions from an interface with a 5 mM salt solution (Reproduced with permission from Ref.[12]). (D) The salt concentrations extracted from the time-resolved force data as a function of time (Reproduced with permission from Ref.[12]). (E) Dynamic interaction forces between tetradecane droplets in three different approaching velocities (Reproduced with permission from Ref.[57]).

In recent years, Chen et al. have conducted many research studies, focusing on the measurements and analysis of interactions between micron-sized droplets[57-59]. In 2018[57], they measured the interaction forces between tetradecane droplets in different concentrations of Sodium dodecyl sulfate (SDS) and NaCl solutions. They found that the droplets coated SDS are negatively charged and the EDL force gradually decreases with the increase of NaCl concentration. Additionally, they observed that absorption amount of the surfactant at the oil-water interface increases with the increase of SDS concentration. In these experiments it was possible to observe the "hydrodynamic suction effect" when the approaching velocity is increased, as shown in Figure 7E. The deformation ratio of emulsion droplets with a diameter of 5 μm is calculated, which means that almost no deformation occurs in the measurements.

In 2019 [59], the same authors found that the tetradecane coated non ionic surfactant FS-30 is negatively charged even though no ionic species were present in the system. Additionally, the screening effect of $Ca^{2+}$, and $Ba^{2+}$ on the EDL between droplets is stronger than that of $Na^+$, which can be explained by the DLVO. In 2020[60], they established the quantitative relationship between the force and the separation distance between droplets using OTs and compared the measurement differences between AFM and OTs. Additionally, a numerical model has been demonstrated to calculate the repulsive pressure from the force curve. The repulsive pressure has the same expression for different sizes of droplets, as it is only a function of the interface separation distance. This model enabled to quantify the measured force between two micron-sized oil droplets coated with polymers and to better understand the interaction mechanism.

A great deal of work has reported how micelles can induce depletion attraction between two colloids. However, the effect of different micelles on the depletion attraction between two emulsion droplets has been rarely reported. Liu et al[61] explored the effect of different micelles on the depletion between two soft surfaces using OTs in 2022. Attractive forces between two like-charged emulsions could be measured. However, for nonionic surfactants, the attractive force between O/W emulsion droplets couldn't be measured even at the CMC of surfactant concentration. The results can explain how surfactant micelles would cause flocculation of emulsions by measuring depletion attraction force between a couple of emulsion droplets in situ. Moreover, it can be used to prepare stable emulsions by adjusting the types and concentration of surfactants.

Another factor affecting emulsion stability is pH, which especially plays a vital role in emulsions for nutrient and drug delivery applications, such as with oleic acid. At pH below 6.5, oleic acid forms oil-like structures, while at higher pH values, it forms oil-in-water emulsions with complex internal nanostructures. Oleic acid is mostly known for usage in common oils but also shows potential for usage in drug delivery systems due to its response to pH[62]. Manca et al. combined a custom-built platform with optical tweezers, polarized optical video microscopy, microfluidics, and small-angle X-ray scattering to investigate the specific mechanisms behind this pH response and the structural changes and interactions among oleic acid molecules. Results showed that depending on the pH, oleic acid molecules go through different phases such as multilamellar vesicles, bicontinuous cubic structures, and hexagonal structures, while also exhibiting self-rotation due to changes in surface tension. For investigation of the interactions between oleic acid particles, the same authors also used double trap OTs. This highlighted that the force of roughly 100 nN applied by the optical tweezers was not strong enough to cause the particles to coalesce, or merge together even at pH as low as 4.0. The same customized setup was also further used by the authors to gain insight into pH-triggered colloidal transformations that play a vital role in e.g. human lipid digestion and drug delivery systems[63]. The authors investigated triolein digestion at single particle level by positioning a digesting triolein droplet inside a microfluidic chip via holographic OTs. Via the chip, pH and pancreatic lipase (an enzyme involved in fat digestion) levels were controlled, while microscopy and small-angle X-ray scattering were used to observe changes in morphology and structure of triolein.

Up to date, emulsion stability has been mainly investigated focusing on the interface properties of single emulsifier, such as ionic surfactants, nonionic surfactants, and polymers. However, emulsions in real life are often stabilized by compound/mixed emulsifiers such as polysaccharide-protein complexes and small molecule surfactant-protein combinations. The interfacial properties of compound/mixed emulsifiers at the oil–water interface that influence the stability of emulsions have not been investigated using OTs. The difficulty lies in the complex components of emulsions, for example, it is challenging to distinguish the effects of multiple factors on emulsion stability, such as the interfacial properties of adsorbed layers and the conformation of emulsifiers under various environmental conditions. In order to study such emulsions with OTs, it is necessary to establish a robust experimental platform and design experiments to distinguish single variables.

*Aggregation and coalescence of emulsion droplets*

The mechanisms of emulsion aggregation and coalescence are particularly important in the food industry. In food, emulsions can produce different effects[64-66]. On the one hand, for food products such as sauces and milk products, aggregation and coalescence should be avoided to extend shelf life and ensure quality and consumer satisfactions. On the other hand, for products such as ice cream, whipped cream or butter, partial coalescence is required to ensure correct structure formation for the desired sensory properties in the sample preparation protocol[67]. While partial coalescence in food emulsions has been widely studied, the mechanism of stabilisation of different partially coalesced states has not been fully understood.

Recently, Mitsunobu et al [68] examined the coalescence of oil droplets stabilized by a surfactant or a hydrophilic polymer using OTs. They observed that droplets could not coalesce at room temperature in spite of the type of emulsifier. In contrast, the coalescence of droplets stabilized by the neutral hydrophilic polymer polyethylene glycol (PEG) was achieved at a temperature higher than 30°C. However, the droplets with ionic surfactants cetyltrimethylammonium bromide (CTAB) or SDS did not coalesce even at high temperature due to their electrostatic repulsion.

The solid content of viscoelastic emulsion droplets can influence their tendency to aggregate and their following coalescence behavior. The balance between the drive to reduce surface tension and the straining of an internal viscoelastic network can create a large number of stable partially-coalesced states[64]. Otazo et al[13] studied the aggregation and subsequent partial coalescence of micro-sized anhydrous milk fat (AMF) droplets by combining OT and a temperature-cycling regime. AMF was chosen to prepare droplets to ensure the presence of crystals in the emulsion. They utilized OTs to make two partly crystalline droplets approach until the distance between them was smaller than the size of the protruding part of the crystal. They used a temperature-cycling regime to adjust the amount of fat crystal in the droplets, which allowed two approaching droplets to gradually merge and take on a spherical shape driven by the Laplace pressure. The use of OTs allows for real-time observation of the aggregation and coalescence processes of partially crystalline emulsion droplets at the microscale. This provides a detailed understanding of the dynamic interactions between droplets and quantitative data on the forces and temperature that lead to aggregation and coalescence. The experimental scheme is shown in Figure 8A. Figure 8B shows arrested coalescence at different temperatures.

Droplet coalescence is also affected by concentrations of specific chemical solutions. This was exemplified by Wen et al. who, via their scanning OTs system, controlled coalescence and splitting of microreactors in femtoliter/picolitre droplets[69]. Increasing ion concentration or exciting fluorescence caused oil droplets to coalesce, either due to the attraction of oppositely charged ions on the droplet surface or weakening of order of the oil molecule arrangement. By addition of an emulsifier and fluorophores into their liquid medium, Wen et al. could also split and stretch oil droplets via excitation of the fluorophores and OT forces.

Aarøen et al [70] investigates how approach velocity affects the likelihood and mechanism of coalescence, revealing the conditions under which droplets are more likely to merge. The depletion force between pairs of droplets was measured based on retract-extend measurements using OTs, which was used to avoid insufficient or excessive contact. Figure 8C-D shows the relationship between force and time during the droplet retract–extend cycles obtained by OTs, top picture shows insufficient contact between the droplets, where depletion force will not be observed during two droplets approaching. Middle picture shows an excessive contact, causing coalescence of two droplets during the pause [2] (in Figure 8D insets), bottom picture shows sufficient contact, where the depletion force was high enough to rearrange two droplets in one trap, and coalescence occurs during [6] (in Figure 8D insets). The coalescence time was defined as the time period from the first encounter between the two droplets until their rupture, as shown in Figure 8D. This is crucial for controlling emulsion properties, especially in processes where maintaining or breaking emulsions is necessary. Understanding the transient behavior of droplets as they approach each other at different velocities provides deeper insights into the stability and dynamics of emulsions over time.

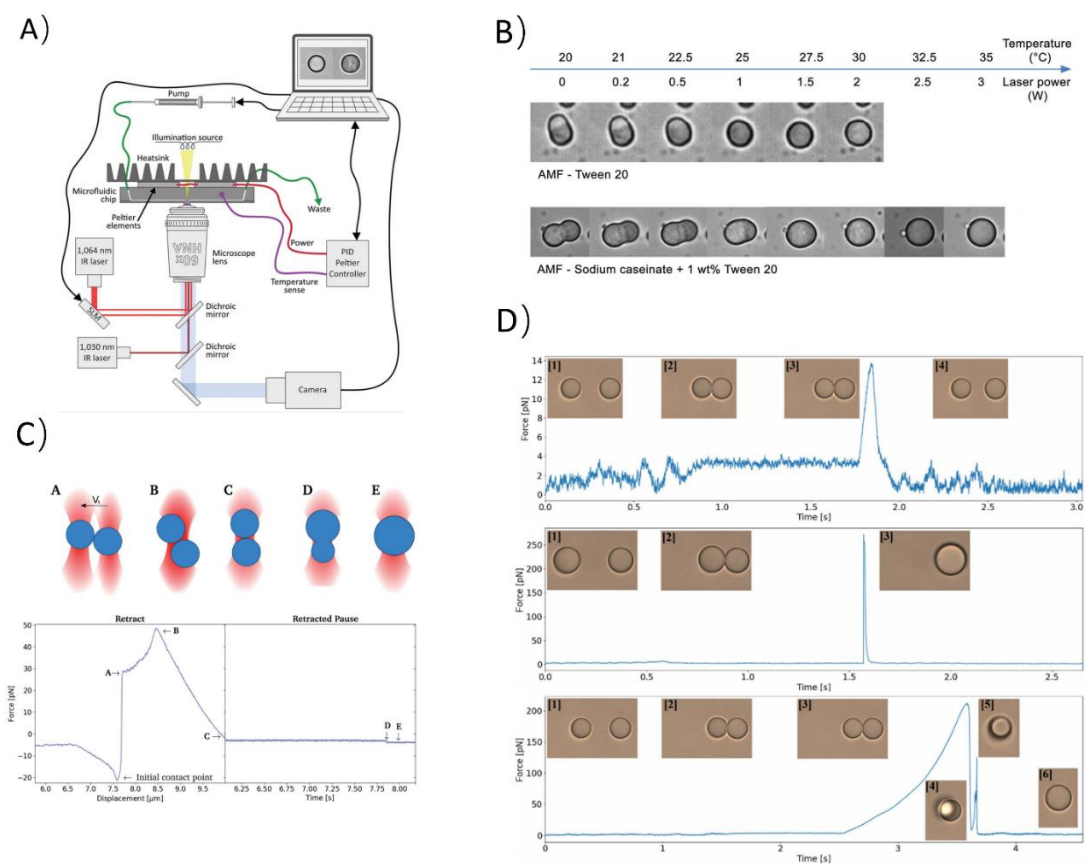

**Figure 8** illustrates (A) Peltier PID controller module fitted to an optical microscope with camera and laser tweezers (Reproduced with permission from Ref.[13]). (B) Arrested coalescence at different temperatures by heating droplets with the HPL at different output powers (Reproduced with permission from Ref.[13]). (C) Trap force versus time curves showing droplet behaviour during retract–extend cycles obtained using OTs (Reproduced with permission from Ref.[70]). (D) Top: insufficient contact between the droplets. Middle: excessive contact between the droplets. Bottom: sufficient contact between the droplets (Reproduced with permission from Ref.[70]).

*Switchable behavior of responsive emulsions*

In some applications, stable emulsions are only temporarily preferred in a certain stage and followed by a controlled demulsification process, which have attracted widespread research interest in various industrial fields including drug delivery, oil transport, and fossil fuel production.

In recent years, switchable or stimuli-responsive emulsions[71] were demonstrated and reversible switch between "emulsification" and "demulsification" by external stimuli or triggers (such as pH[72, 73], temperature[74, 75], light irradiation[76], redox[77, 78], magnetic field[79], $CO_2/N_2$[80, 81], or multiple stimuli[82]) were reported. The core process of these systems is the switchable behavior between emulsification and demulsification, which is inseparable from the stability and instability of the emulsions. The quantitative measurements and analysis of interactions between a pair of switchable emulsion droplets are urgently desirable.

Switchable surface-active colloid particles are crucial for the preparation of switchable Pickering emulsions (PE)[82-85]. Researchers can explore particle dynamics, assembly processes, and phase transitions as a basis for extending OTs techniques to the study of emulsion droplets and other complex systems. In general, the initial colloidal particles are usually so hydrophilic and surface-inactive that they couldn't prepare stable PE. To solve these limitations, many previous studies have provided effective methods for partially hydrophobized colloidal particles by adsorbing switchable surfactants with opposite charges, enabling the preparation of switchable PE through certain triggers. Chen et al [86] developed a novel approach to measure the interaction forces between a couple of switchable surface-active colloid particles in situ using OTs. They prepared switchable surface-active silica particles by partially hydrophobizing commercially available inorganic silica particles in water using the common cationic surfactant CTAB. Furthermore, the surface-active form can be converted to the surface-inactive form at room temperature by using the conventional anionic surfactant SDS. Figure 9A shows a diagrammatic sketch of force measurement between switchable surface-active silica particles. Figure 9B showed interaction forces of switchable surface-active colloid particles.

Bauer et al [87] combined the concepts of engineered emulsions with the advantages of the microfluidic methods. It is possible to generate monodisperse, functional O/W droplets stabilized by a pH-responsive copolymer surfactant in microfluidic devices. Aggregation and disaggregation driven by inter-droplet hydrogen bonds formed macroscopic structures and dispersed structures, which were controlled by a simple pH trigger. PH-dependent interactions between individual droplets were quantitatively analyzed using OTs.

Chen et al [88] measured the interaction forces between the $CO_2$-responsible switchable behaviors of demulsification and restabilization using OTs and revealed the switchable mechanism. They introduced $CO_2$ and $N_2$ into emulsion droplets and achieved detachment / self-assembly of the switchable surfactant, which caused the desorption and reabsorption of the switchable surfactant from the water−oil interface, leading to the weakening and re-enhancing of the EDL repulsive forces between emulsion droplets. Figure 9C shows schematic diagram of the measurements of interaction force between a couple of individual switchable emulsion droplets and the process of stimulus responsivity of the switchable surfactant by $CO_2/N_2$ trigger. Figure 9D showed the force curve of droplets between the processes of emulsification and demulsification.

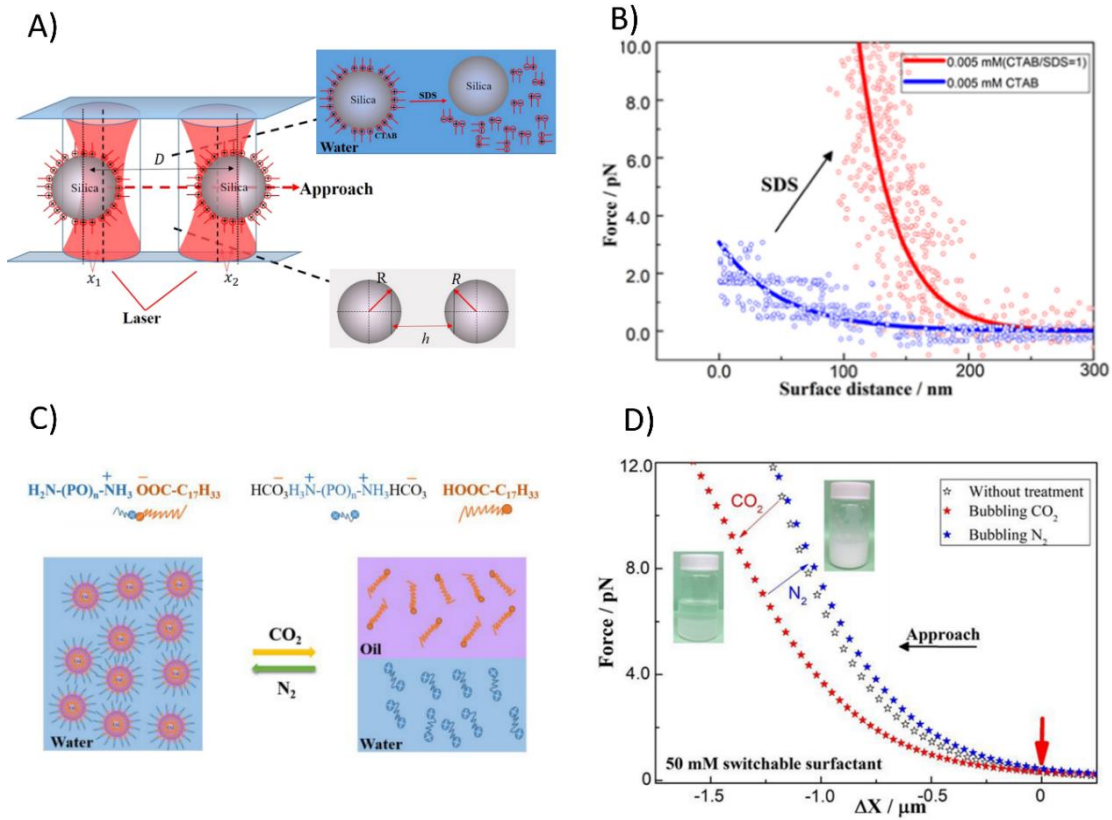

**Figure 9** (A) Diagrammatic sketch of force measurement between switchable surface-active silica particles (Reproduced with permission from Ref.[86]). (B) Interaction forces between two 5.0 μm silica particles in 0.005 mM CTAB solution or with the addition of equimolar SDS molecules (Reproduced with permission from Ref.[86]). (C) Schematic diagram of the measurements of interaction force between a couple of individual switchable emulsion droplets and the process of stimulus responsivity of the switchable surfactant by $CO_2/N_2$ trigger (Reproduced with permission from Ref.[88]). (D) Force curve of 5.0 μm (diameter) tetradecane droplets between the processes of emulsification and demulsification upon bubbling $CO_2$ or $N_2$ alternatively (Reproduced with permission from Ref.[88]).

**Instrumentation**

Previously, the applicability of OTs in relation to emulsions was discussed. OTs can achieve a higher efficiency in certain applications than current widely used methods. For instance, in isolation and separate encapsulation of individual cells, OTs are superior to a statistical approach to encapsulation, and also to commonly used sorting strategies[89, 90]. However, their spread is limited, one possible reason for which may be the complexity of their instrumentation. Another possible limitation is affordability, as OTs require a combination of high-end instruments to set up[91]. In this section, we overview the instrumentation framework of various OT setups, and evaluate their complexity and cost with a view towards the democratization of this technology.

To compare OT setups in terms of cost and complexity, we take essential functional elements, as well as the main cost-drivers (Table 1). These are: 1) objective, 2) number of beam-forming and beam-steering optical components (light sources and detectors are not counted, but the objectives are), 3) light source used for entrapment, and 4) the detector used for imaging[91]. We also compare the application and the object the laser beams entrap. While the number of beam-steering optical components may not be directly comparable due to the diverse applications, it can be indicative of a difference (if there is one) between setups declared as "low-cost" in the literature and setups that are not. Applications in Table 1 are categorized using the ontology introduced in the subsections of this section.

As indicated by Figure 3, open-source/low-cost OT systems have entered publishing very recently, and systems for manipulating emulsions/droplets occurred first in 2020. In terms of instrumentation however, they are similar, and thus we will report on both, starting with systems dedicated to the manipulation of droplets/emulsions. Suwannasopon et al[38] demonstrated a setup for driving NLC droplets (Figure 4). It used a 1064 nm fiber laser operated at 300 mW, which was circularly polarized by a Glan-Thompson polarizer and quarter waveplate (QWP). NLC droplets were held in a glass slide chamber filled with 4.5 μm polystyrene beads, in direct contact with a metalens. A CCD camera was used to image the movement of the beads through a 40x objective. Xu et al[89] presented the EasySort system (Figure 10B) for OT-assisted pool-screening and single-cell isolation (OPSI), that is, capture of individual cells (1-40 μm) and selective encapsulation in nanoliter droplets. The authors claimed >99.7% sorting accuracy with a throughput of 10-20 cells/min. While the sorting throughput is significantly lower than more traditional label-free droplet-based cell sorting methods (~40 droplets/second), the accuracy is considerably higher (~90-95%)[92, 93]. Zhai et al[44] performed super-resolution microscopy by suspending microdroplets above the object under observation using an OT (Figure 5C). Compared to its utility, the experimental setup was fairly simple. An ytterbium fiber laser (IPG Photonics YLR-5-LP) was used with a high NA lens (ZEISS 63x, NA:1.2). The experimental setup of Chen et al.[43] consisted of an inverted fluorescence microscope (Nikon Eclipse Ti), and a scanning optical tweezing system (Aresis Tweez 250si) and was used to manipulate lipid droplets in mature adipose cells.

Interaction force measurements in emulsions is another unique application area of OTs. Several authors in this field used standard optical tweezer instrument setups. For instance, Liu et al[53] measured interaction forces (Figure 6A) between microparticles in an emulsion using the same Aresis Tweez 250si system (Figure 10A) as Chen et al.[43] with no apparent modifications to the optical path. Julie et al[56] used on a Nanotracker (JPK Instruments) mounted on an inverted light microscope (Zeiss Axio Observer A1), and Chen et al.[57] used a Nanotracker 2 (JPK Instruments). Griffiths et al.[12] built their setup (Figure 10C) on the basis of a HOT (Arryx Inc., Holographic Optical Tweezers) system and used two diode laser beams, one movable (1064 nm, 2W) and one stationary (1030 nm, 5W).

Aggregation/coalescence of emulsions is widely researched (Figure 3). Otazo et al[13] also used the Arryx HOT system with the same dual laser beams, albeit with a different set of objectives with a higher magnification (60x Nikon MRD07602, NA 1.2 + auxiliary 1.5x lens resulting in a 90x total). Aarøen et al [70] used a Nanotracker 2 (JPK Instruments) mounted on a Zeiss Axio Observer Inverted optical microscope. The laser used was a TEM00 with a 3W maximum power. Mitsunobu et al[68] also used an inverted microscope (Nikon TE2000-S) as the basis of their experimental setup, and outfitted it with a 532 nm, 3W, Nd:YAG laser (Beamtech Optronics). The laser beam was split by two beam splitters, and the resultant beams were introduced to the microscope objective via dichroic mirrors.

**Figure 10** Instrumentation platforms used for optical tweezing of emulsions show a high

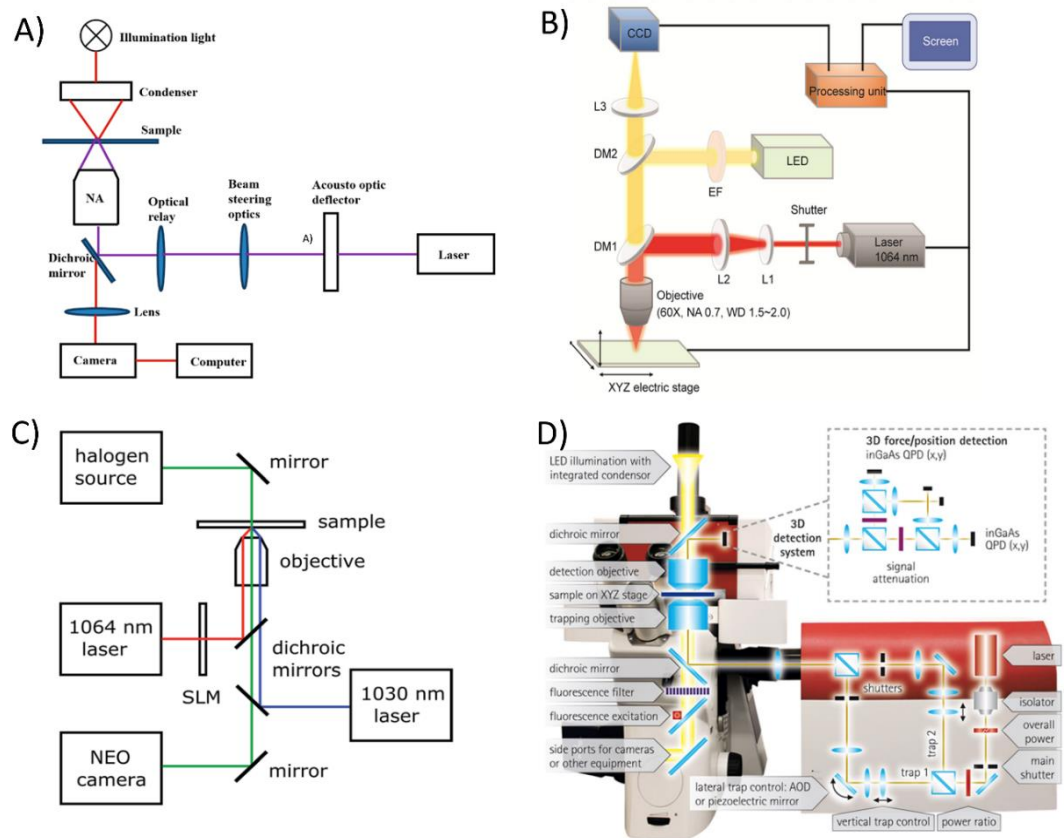

degree of similarity in their optical path with typically minor differences on the level of single components. (A) In fact, more than one group relied on the same standard setup (Aresis Tweez 250si [43, 53], Reproduced with permission from Ref. [53]). (B) Low-cost implementations are so far uncommon and in terms of the optical components, show little to no difference to typical setups (Reproduced with permission from Ref. [89]). (C) A dual-beam setup (one fixed, one movable trap) can be advantageous for interaction force measurements, such as the layout in the Arryx HOT system used by several groups[12] [13] (Reproduced with permission from Ref. [12]). (D) The JPK (now Bruker JPK) Nanotracker is another popular platform used by several groups for various applications including interaction force measurements, droplet aggregation/coalescence studies, and stimuli-responsive emulsion studies. The platform combines an optical tweezer and an inverted microscope [12, 57, 70, 86] (image: JPK NanoTracker 2 | Bruker).

Table 1: Comparison of OT instrumentation setups for the manipulation of emulsions.

| Application group | Reference | Beam-forming/steering optical components (total pieces) | Objective | Tweezer light source | Tweezed object | Imaging detector |
|---|---|---|---|---|---|---|
| Manipulation of emulsion droplets (including manipulation of objects in emulsion droplets) | [38] | 9 | 40x (for imaging) | 1064 nm @ 300 mW (circularly polarized CW fiber laser) | NLC droplets (5CB-CTAB) | Camera (CCD) |
| | [89] | 9 | 60x (Olympus 60x, NA 0.7) | 532 nm @ 50 mW for imaging, 1064 nm laser @ 150 mW for sorting | E. coli, yeast cells (isolating 10-20 cells/min with >99% accuracy) | Camera (CCD) |
| | [44] | Exact number not shown (several 450 nm bandpass filters) | 63x (ZEISS, NA:1.2) | 1064 nm @ 5W (IPG Photonics YLR-5-LP) | Microdroplet | Camera (Manta G-507 monochrome) |
| | [63] | 8 | 100x (Nikon CFI Plan 100 XC W) | 1064 nm (IPG photonics YLR-10-LP) | Triolein droplet | Camera (Nikon DS-Qi2) |
| | [43] | 10 | 60x (Nikon water-immersion, | 1064 nm CW (part of Aresis Instruments Tweez | Mature adipose cells with lipid droplets | Camera (CCD) and spectrometer (SHIS VNIR-520-20-S, |

| | | | NA 1.0) | 250si) | | Wayho Technology, Guangzhou, China) |
|---|---|---|---|---|---|---|
| Emulsion stability studies (primarily interaction force measurement) | [53] | 7 | 60x (Nikon water-immersion, NA 1.0) | 1064 nm @ 5W (part of Aresis Instruments Tweez 250si) | Silica particles in sodium dodecyl benzene sulfonate (SDBS) solution | Camera (CCD) |
| | [12] | 6 | 60x (Nikon Eclipse TE2000-U, NA 1.2) | 1064 nm @ 2 W (diode laser, movable trap) and 1030 nm @ 5W (diode laser, fixed trap) | Silica beads; sodium-caseinate-stabilized soybean oil droplets | Camera (16-bit CMOS, Andor NEO) |
| | [62] | 8 | 100x (Nikon, CFI Plan 100X C W water-immersion, NA 1.10) | 1064 nm laser (Nd:YAG crystal laser, IPG photonics, model YLR-10-LP) | Oleic acid droplets | Camera (CCD, Teledyne FLIR LLC Flea3 FL3-U3-13E4M) |
| Aggregation and coalescence of emulsion droplets | [13] | 6 | 90x (60x Nikon MRD | 1064 nm @ 2 W (diode laser, movable | Micro-sized anhydrous milk fat (AMF) | Camera (16-bit CMOS, Andor NEO) |

| | | | 07602, NA 1.2 + auxiliary 1.5x lens) | trap) and 1030 nm @ 5W (diode laser, fixed trap) | droplets | |
| --- | --- | --- | --- | --- | --- | --- |
| [68] | | 9 | 40x (NA 0.6) | 532 nm @ 3W (Beamtech Optronics Nd:YAG laser) | Oil droplets | Camera (Watec WAT-221S CCD) |

Low-cost and open-source are new concepts in optical tweezers, having started around 2019, with ~5-10 new publications per year in the last 5 years (Figure 3). Some fields, such as droplet-based single-cell isolation, could significantly advance by means of emulsion OT technology. In these fields, more affordable instrumentation could provide a significant boost to application development.

At present however, some limitations to this sub-field exist: only particular components, such as software used for analysis are open-source, and are not self-developed, whereas self-developed hardware systems declared by the authors as "low-cost" typically have no cost calculation included for comparison. They also show little to no difference in terms of instrumentation, as compared to setups not reported as low-cost. Finally, publishing rate seems to decrease over time. There may be some technical challenges limiting increased growth in this research area.

**Optical Tweezers for emulsions – limitation in the experimental method**

Although the use of micropipette and OT in single trap configurations was reported in the study of colloids[49, 51], a dual-laser OTs is typically used to study couples of emulsion droplets. In a dual-laser configuration, one of the beads/emulsions (held in a steerable trap) is stepped towards the other beads/emulsions (held in a fixed trap). In order to exclude the effect of hydrodynamic force, the approaching velocity is usually adjusted below 1.0 μm/s [57], and when the two droplets get close to each other and start to interact, the force between them can be calculated [94].

During sample preparation, the original emulsion solutions should be diluted to obtain a proper density of droplets in the sample chamber. Emulsions tend to adhere to sample chamber making the trapping experiment difficult. To limit the adhesion between droplets and sample chamber, the surface treatments should be used. For instance, Julie et al. [56] reported on the use of 1 mg/ml BSA solution to coat surface of the cover glass for 60 min, while Murakami et al.[52] used a coating of (3-(2-aminoethyl)aminopropyl)trimethoxysilane to prevent the adsorption of particles onto the fluidic chamber surface. However, the identification of a more standard protocol to prevent adhesion would be preferable.

Another major challenge in using OTs for the study of emulsions is data reproducibility. In particular, the interaction of different droplets in different OT experiments can be significantly different due to differences in droplet size, and the adsorption of the emulsifier at the oil-water interface. The resolution of the light microscope does not always allow a precise determination of the contact point between droplet surfaces. In order to improve data reliability and reproducibility it is critical to obtain samples as chemically homogeneous as possible prior to emulsion experiments. For example, using a droplet generation device in microfluidic to obtain uniform droplets. Researchers did important efforts to obtain a measurement interval under a certain condition, such as the same type of emulsifiers, droplet size, approach velocity and solution properties, as mentioned by Julie et al.[56] and Ola et al.[70]. Interestingly, OTs combined with microfluidic channels can achieve interaction measurement between the same pair of emulsion droplets in different environmental conditions in situ[12, 53, 86].

OT-technology is not suitable for the trapping and analysis of droplets in W/O emulsion systems as the refractive index of the water phase is usually lower than oil phase. This limits the direct use of OT-s in droplet microfluidics where W/O droplets are increasingly being applied: e.g. in diagnostics[6], single cell analysis[95], screening for novel drugs[7] and enzymes[96]. However, one can easily envision OT-s being used together with droplet microfluidics for trapping and analyzing cells of interest for downstream encapsulation into W/O droplets and further analysis (e.g. genomics). Biological cells usually have higher refractive index than their surrounding water-based medium[97]. OT-s have been shown already be effective in sorting cells of different sizes into water-droplets for further genomic analysis in low-throughput settings[89]. The open challenge then remains to develop further the OT technology in combination with droplet microfluidics to enable such analysis in high-throughput. So far, various precise, flexible and high-throughput manipulation techniques have been developed. Optoelectronic tweezers (OET)[98] is an advanced technique combining light stimuli with electric field together by utilizing the photoconductive effect of semiconductor materials, which can be used to manipulate water droplets in water. Additionally, the use of donut beams to trap low-index particles can be used to trap and manipulate oil in water droplet. Gahagan et al[99] already reported the low-index particle trapping in 1999. Garbin et al[100] reported on the use of donut beams to trap UCA (ultrasound contrast agent) micro bubbles, thus it is possible to manipulate oil in water droplets through the use of donut beams.

**Outlook**

In conclusion, OTs have emerged as a potential tool for the study of emulsions thanks to its high spatial and temporal resolution and high sensitivity in measuring forces. Moreover, OTs enable the suspension of the emulsion in the specified position in the liquid and to control the environmental conditions of the emulsion.

To date, OT can be combined with emulsion droplet to form a special functional optical device, such as ideal optical motors and droplet-assisted imaging system. Additionally, the stability mechanism of emulsion stabilized by single emulsifier has been studied by measuring the force and displacement between two dispersed droplets by OT, even utilizing an instability mechanism to control aggregation and coalescence of emulsions. Moreover, switchable behavior of pH-responsive and $CO_2$-responsive emulsions has been investigated.

In the future, OT will be a promising tool to explore the interfacial properties of compound/mixed emulsifiers at the oil–water interface, which is the basis for studying complex emulsions. The study of stimuli-responsive emulsions by OTs is a new interesting topic, response mechanisms of other stimuli-responsive emulsions is expected to be studied by different triggers, such as temperature, light and magnetic fields.

**Acknowledgments**

We are grateful for continued financial support from National Natural Science Foundation of China (No. 22202167), National Key Research and Development Project of China (No. 2023YFF0613603), Provincial Science and Technology Plan Project: Micro and Nano Preparation and Photoelectronic Detection (No. 03014/226063), the European Union Program HORIZON-Pathfinder-Open:3D-BRICKS, grant Agreement 101099125.